\documentclass{PoS}

\usepackage{bbm,latexsym,amssymb,amsmath,amsbsy,epsfig}

\usepackage{stmaryrd,xcolor}
\usepackage{amsfonts,graphicx,epsfig,float,ulem}
\usepackage{bbm}
\usepackage{color}
\usepackage{array}

\newcommand{\be}{\begin{equation}}
\newcommand{\ee}{\end{equation}}
\newcommand{\ba}{\begin{eqnarray}}
\newcommand{\ea}{\end{eqnarray}}
\newcommand{\bs}{\begin{subequations}}
\newcommand{\es}{\end{subequations}}

\makeatletter
\newcommand{\thickhline}{%
    \noalign {\ifnum 0=`}\fi \hrule height 1.5pt
    \futurelet \reserved@a \@xhline
}
\newcolumntype{"}{@{\hskip\tabcolsep\vrule width 1.5pt\hskip\tabcolsep}}
\makeatother

\title{A minimal seesaw model with mu-tau symmetry}

\ShortTitle{A minimal seesaw model with mu-tau symmetry}

\author{\speaker{Darius Jur\v{c}iukonis}$^{,a}$, Thomas Gajdosik$^{a,b}$ and Andrius Juodagalvis$^a$\\
   \llap{$^a$} Vilnius University, Institute of Theoretical Physics and Astronomy\\
   \llap{$^b$}  Vilnius University, Physics Faculty\\
   E-mail: \email{Darius.Jurciukonis@cern.ch},
   \email{Thomas.Gajdosik@cern.ch},
   \email{Andrius.Juodagalvis@cern.ch}}

\abstract{We analyse a flavour model for a lepton sector which is
  based on type I seesaw mechanism, a $\mathbbm{Z}_2$ symmetry for
  lepton flavours, a $\mu$--$\tau$ interchange symmetry and a $CP$
  symmetry. This model fits well the data of neutrino mass squared
  differences and oscillation angles. The model predicts an overall
  neutrino mass scale for normal and inverted neutrino mass hierarchy
  and the effective mass $m_{\beta\!\beta}$, which is used in
  the neutrinoless double beta decay.}

 \FullConference{The European Physical Society Conference on High Energy Physics\\
                  22-29 July 2015\\
                  Vienna, Austria}

\begin{document}


\section{The model}

Using the ideas of ref.~\cite{Ferreira:2013zqa} 
we modify the model of ref.~\cite{Grimus:2001ex}: 
we get neutrino masses by the type~I seesaw 
mechanism and restrict the Lagrangian with a $\mathbbm{Z}_2$ symmetry for 
each lepton flavour and a $CP$ symmetry incorporating the interchange of 
two lepton flavours.

We take the lepton sector with three right-handed neutrinos and
impose the three family lepton numbers $L_e$,
$L_\mu$,
and $L_\tau$.
Therefore,
the lepton sector comprises the multiplets $D_{eL}$,
$D_{\mu L}$,
$D_{\tau L}$,
$e_R$,
$\mu_R$,
$\tau_R$,
$\nu_{eR}$,
$\nu_{\mu R}$,
and $\nu_{\tau R}$.
The scalar sector of the model includes three doublets:
\be
\phi_k = \left( \begin{array}{c} \varphi_k^+ \\*[0.5mm] \varphi_k^0
\end{array} \right),
\quad
\tilde \phi_k
= \left( \begin{array}{c} {\varphi_k^0}^\ast \\*[0.5mm] - \varphi_k^-
\end{array} \right), \quad k = 1, 2, 3.
\ee
In addition to the family-lepton-number symmetries we have
three symmetries in the model. The first one ensures that 
only $\phi_3$ has Yukawa couplings
to the electron family:
\be
\mathbbm{Z}_2^{(e)}: \quad D_{eL} \to - D_{eL},\ \phi_3 \to - \phi_3.
\ee
The second symmetry is the $\mu$--$\tau$ interchange symmetry
\be
\mathbbm{Z}_2^{(\mathrm{int})}:\quad
D_{\mu L} \leftrightarrow D_{\tau L},\
\mu_R \leftrightarrow \tau_R,\
\nu_{\mu R} \leftrightarrow \nu_{\tau R},\
\phi_2 \to - \phi_2.
\ee
Notice that $\phi_2$ changes sign under
$\mathbbm{Z}_2^{(\mathrm{int})}$.
If $\varphi_2^0$ acquires a vacuum expectation value (vev) $v_2$,
the $\mu$--$\tau$ interchange symmetry gets broken.
The third symmetry is the $CP$ symmetry
\be
\label{cpcp}
\begin{array}{rcl}
& &
\left\{
\begin{array}{rcl}
D_{eL} (x) &\to& \gamma_0 C \bar D_{eL}^T (\bar x)
\\*[1mm]
D_{\mu L} (x) &\to& \gamma_0 C \bar D_{\tau L}^T (\bar x)
\\*[1mm]
D_{\tau L} (x) &\to& \gamma_0 C \bar D_{\mu L}^T (\bar x)
\end{array}
\right.,
\quad
\left\{
\begin{array}{rcl}
e_R (x) &\to& \gamma_0 C \bar e_R^T (\bar x)
\\*[1mm]
\mu_R (x) &\to& \gamma_0 C \bar \tau_R^T (\bar x)
\\*[1mm]
\tau_R (x) &\to& \gamma_0 C \bar \mu_R^T (\bar x)
\end{array}
\right.,
\\*[-2mm]
CP: & &
\\*[-2mm]
& &
\left\{
\begin{array}{rcl}
\nu_{eR} (x) &\to& \gamma_0 C \bar \nu_{eR}^T (\bar x)
\\*[1mm]
\nu_{\mu R} (x) &\to& \gamma_0 C \bar \nu_{\tau R}^T (\bar x)
\\*[1mm]
\nu_{\tau R} (x) &\to& \gamma_0 C \bar \nu_{\mu R}^T (\bar x)
\end{array}
\right.,
\quad
\left\{
\begin{array}{rcl}
\phi_1 (x) &\to& \phi_1^\ast (\bar x) \\
\phi_2 (x) &\to& - \phi_2^\ast (\bar x) \\
\phi_3 (x) &\to& \phi_3^\ast (\bar x)
\end{array}
\right.,
\end{array}
\ee
where $x = \left( t,\, \vec{r} \right)$
and $\bar x = \left( t,\, - \vec{r} \right)$.
Notice that $CP$ interchanges the $\mu$ and $\tau$ lepton flavours
and that $\phi_2$ changes sign under $CP$.

Majorana masses of the right-handed neutrinos are generated at the 
(very high) seesaw scale and are given by
\ba
\mathcal{L}_\mathrm{Majorana} =
- \frac{1}{2}
\left( \begin{array}{ccc}
\bar \nu_{eR}, &
\bar \nu_{\mu R}, &
\bar \nu_{\tau R}
\end{array} \right)
M_R\, C
\left( \begin{array}{c}
\bar \nu_{eR}^T \\
\bar \nu_{\mu R}^T \\
\bar \nu_{\tau R}^T
\end{array} \right) +  \mathrm{H.c.}
\ea
where $M_R$ is a $3 \times 3$ symmetric matrix.
These mass terms violate the family lepton numbers softly,
but they are not allowed to violate
neither $\mathbbm{Z}_2^{(\mathrm{int})}$ nor $CP$.
$M_R$ has a typical $\mu$--$\tau$ symmetric form with real numbers.

The Yukawa couplings have the dimension four and must conserve 
the family lepton numbers. They are real because of the $CP$ symmetry
and together with the vevs produce
the Dirac mass terms of the leptons
\be
\mathcal{L}_\mathrm{Dirac} =
-
\left( \begin{array}{ccc}
\bar e_L, & \bar \mu_L, & \bar \tau_L \end{array} \right)
M_\ell
\left( \begin{array}{c} e_R \\ \mu_R \\ \tau_R \end{array} \right)
-
\left( \begin{array}{ccc}
\bar \nu_{eR}, & \bar \nu_{\mu R}, &
\bar \nu_{\tau R} \end{array} \right)\,
M_D
\left( \begin{array}{c}
\nu_{eL} \\ \nu_{\mu L} \\ \nu_{\tau L}
\end{array} \right)
+ \mathrm{H.c.}
\ee
In this model $M_\ell = \mathrm{diag} \left( x_e,\, x_\mu,\, x_\tau \right)$,
and $M_D = \mathrm{diag} \left( a_e^\ast,\, a_\mu^\ast,\, a_\tau^\ast \right)$.
The charged-lepton masses are
$m_\alpha = \left| x_\alpha \right|$, where $\alpha = e, \mu, \tau$.

Using the fact that $M_R$ is at a much higher scale than $M_D$,
we may use the see-saw formula to obtain
the effective light-neutrino mass terms
\be
\mathcal{L}_{\mathrm{light}\, \nu\, \mathrm{mass}} =
\frac{1}{2} \sum_{\alpha, \beta = e, \mu, \tau}
\nu_{\alpha L}^T \left( M_\nu \right)_{\alpha \beta} \nu_{\beta L}
+ \mathrm{H.c.},
\ee
where
\be
M_\nu \approx - M_D^T M_R^{-1} M_D.
\label{bviut}
\ee
\begin{figure}[t]
\begin{center}
\includegraphics[scale=0.88]{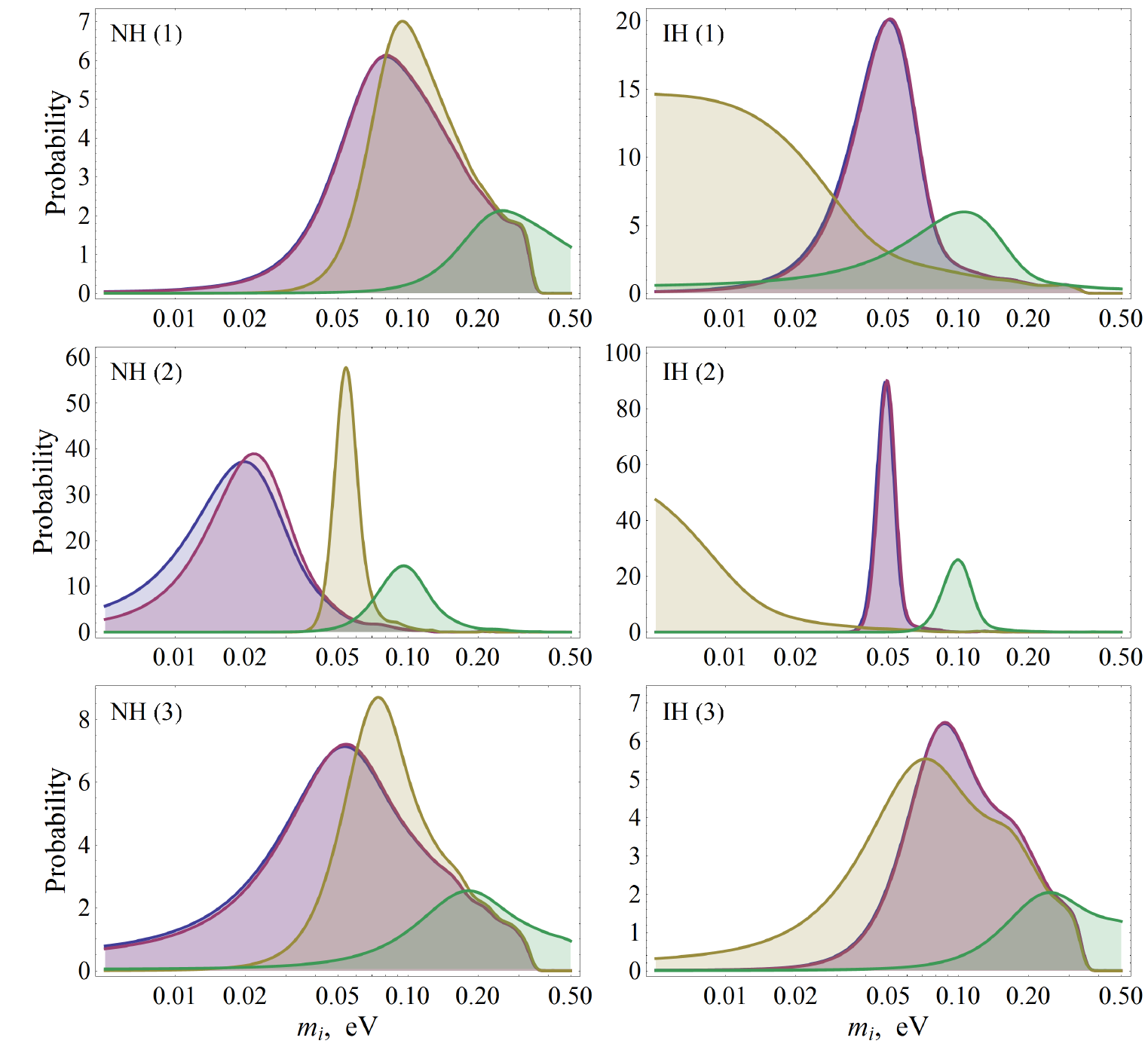}
\end{center}
\vspace{-0.7cm}
\caption{Distributions of the neutrino masses for the cases 1, 2, and
  3 of table~1. Normal and inverted hierarchies of the
  neutrino masses (NH and IH) are analysed
  and shown on the left and right, respectively.
  The curves of blue, purple, and yellow colors represent
  the masses of the light neutrinos, $m_1$, $m_2$, and $m_3$,
  respectively. The green color indicates the distribution of $\sum
  m_i$.}
\label{picture1}
\end{figure}%
The symmetries of the model,
$\mathbbm{Z}_2^{(\mathrm{int})}$ and $CP$,
lead to the following constraints on the matrix elements of $M_\nu$:
\bs
\begin{align}
& \left| M_{\nu,22} M_{\nu,13}^2 \right| = \left| M_{\nu,33} M_{\nu,12}^2 \right|,
\\
& \arg{\left( M_{\nu,11}^\ast M_{\nu,22}^\ast M_{\nu,12}^2 \right)}
= \arg{\left( M_{\nu,11}^\ast M_{\nu,33}^\ast M_{\nu,13}^2 \right)} =
                          0\ \mathrm{or}\ \pi,
  \hspace{0.5cm} \arg{\left( M_{\nu,22}^\ast M_{\nu,33}^\ast M_{\nu,23}^2 \right)} = 0.
\label{m2}
\end{align}
\es
This corresponds to one condition on the moduli
and three conditions on the phases of the neutrino masses and mixings.
The conditions~(\ref{m2}) mean that
$M_\nu$ is real.


\section{Numerical analysis}

We write the neutrino mass matrix as
\be
M_\nu = U^\ast\, \mathrm{diag} \left( m_1,\, m_2 e^{- i \alpha_{21}},\,
m_3 e^{- i \alpha_{31}} \right) U^\dagger,
\ee
where the $m_k$ ($k = 1, 2, 3$) are the neutrino masses, which are
real, and $\alpha_{21}$ and $\alpha_{31}$ are the Majorana phases. The
unitary matrix $U$ is parameterized like the PMNS mixing
matrix~\cite{Agashe:2014kda} which contains three oscillation angles
$\theta_{12},\theta_{13},\theta_{23}$ and one $CP$ violating Dirac phase
$\delta$.

Since the neutrino mass matrix $M_\nu$ is real, the phases
$\alpha_{21}$, $\alpha_{31}$, and $\delta$ may be either $0$ or $\pi$.
We make eight separate numerical fits to the experimental
data~\cite{Capozzi:2013csa},
according to the distinct
values of the phases, listed in table~\ref{tabl1}.
Both normal and inverted hierarchies of the light neutrino masses are
considered.
If the Majorana phases fulfil the condition
$e^{i\alpha_{23}}=e^{i\alpha_{31}}=-1$ (cases 4 and 8), acceptable
fits cannot not be obtained,
but other cases yield good results.

We use the $3 \sigma$ intervals for the
experimental values of
$\theta_{ij}$,
for mass-squared difference $\Delta m_{21}^2 = m_2^2 - m_1^2$,
and for $\left| \Delta m^2 \right| =
\left| m_3^2 - \left. \left( m_2^2 + m_1^2 \right)\,
\right/ 2 \right|$
as given in ref.~\cite{Capozzi:2013csa}.
We impose an upper limit on the sum of the neutrino masses: 
$m_1 + m_2 + m_3 \le 1\, \mathrm{eV}$.

\begin{table}[tb]
\begin{center}
\begin{tabular}{c"c|c|c|c|c|c|c|c}
case & 1 & 2 & 3 & 4 &  5 & 6 & 7 & 8
\\ \thickhline
$e^{i \alpha_{21}}$ & $+1$ & $+1$ & $-1$ & $-1$ & $+1$ & $+1$ & $-1$ & $-1$  \\
$e^{i \alpha_{31}}$ & $+1$ & $-1$ & $+1$ & $-1$ & $+1$ & $-1$ & $+1$ & $-1$  \\
$e^{i \delta}$      & $+1$ & $+1$ & $+1$ & $+1$ & $-1$ & $-1$ & $-1$ & $-1$  \\
\end{tabular}
\caption{Values of $e^{i \alpha_{21}}$, $e^{i \alpha_{31}}$,
and $e^{i \delta}$ for different cases of fits.\label{tabl1}}
\end{center}
\end{table}

\begin{figure}[t]
\begin{center}
\includegraphics[scale=0.88]{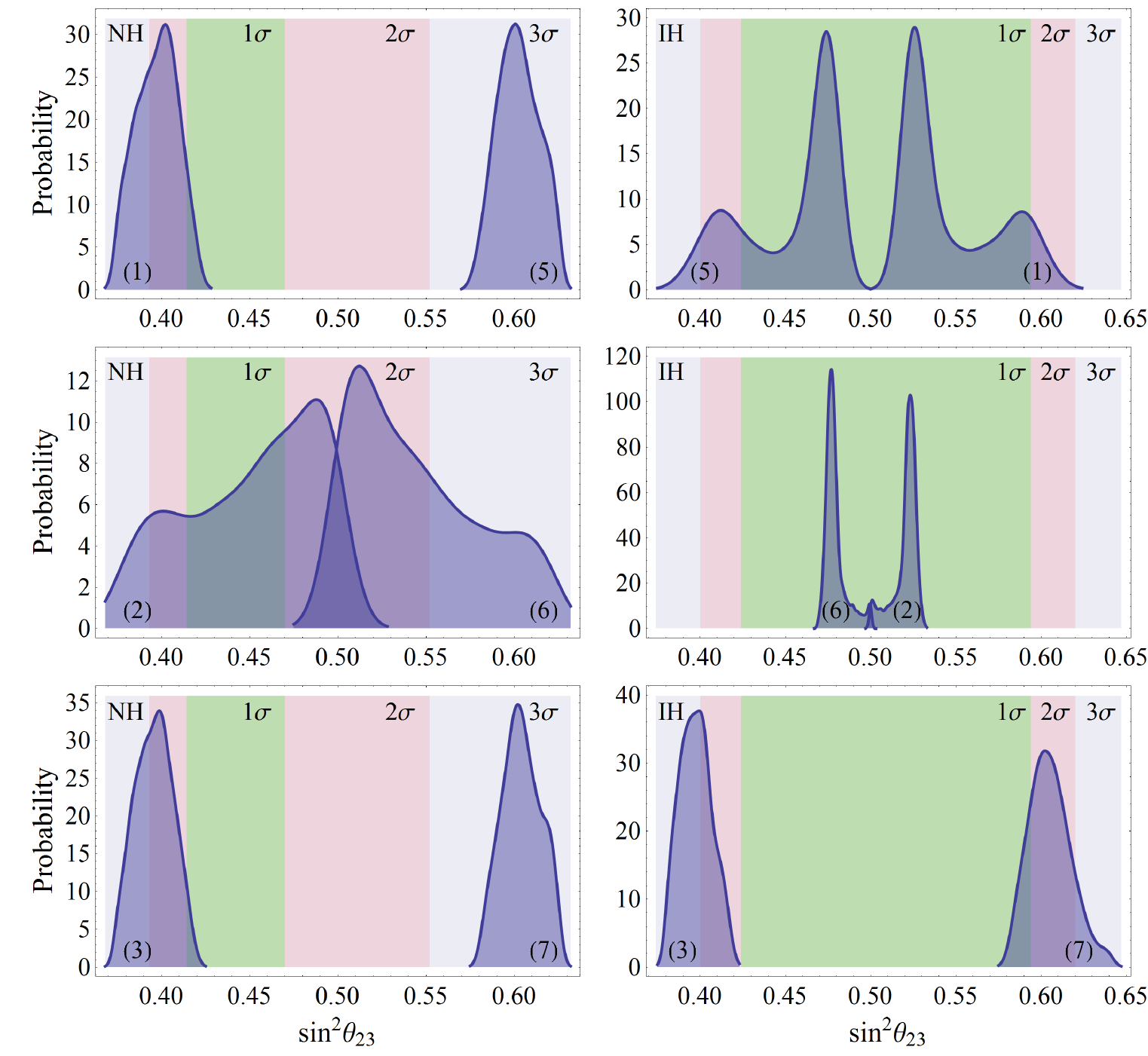}
\end{center}
\vspace{-0.7cm}
\caption{Distributions of the oscillation angle $\sin^2\theta_{23}$
  for different cases of table~1.
  The case is indicated by a number in parantheses.
  Normal and inverted hierarchies
  of the neutrino masses (NH and IH) are shown on the left and right,
  respectively.
  The blue, red, and green colors of the
  background represent the $3\sigma$, $2\sigma$, and $1\sigma$
  intervals of the experimental data.
\label{picture2}
}
\end{figure}

\begin{figure}[t]
\begin{center}
\includegraphics[scale=0.88]{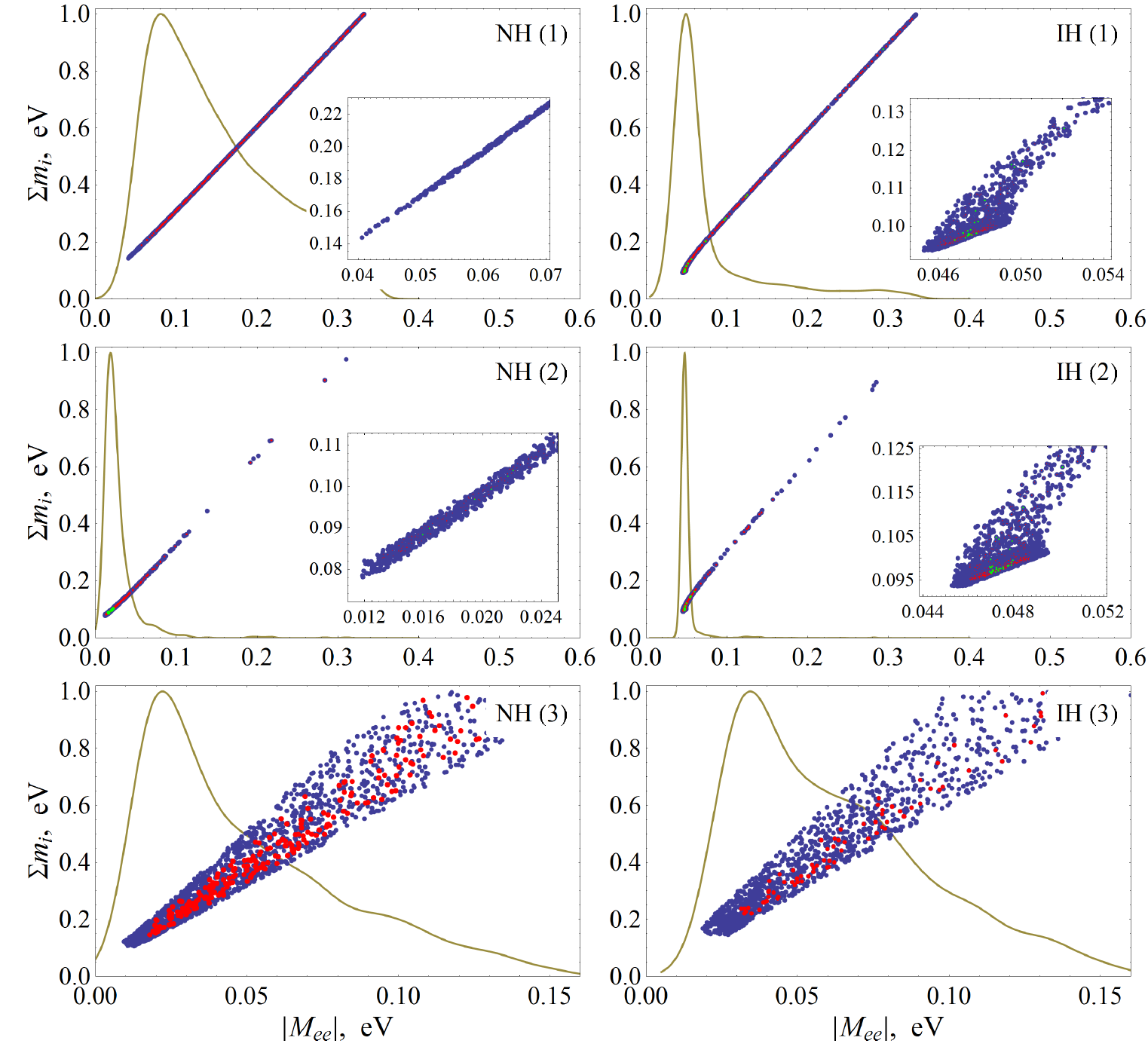}
\end{center}
\vspace{-0.7cm}
\caption{Plots for the sum of the neutrino masses $\sum m_i$ {\it
    vs}.\ $m_{\beta\!\beta}=|M_{ee}|$ for the cases 1, 2, and 3 of
  table~1.
  Normal
  and inverted hierarchies of the neutrino masses (NH and IH) are analysed.
  The blue, red,
  and green colors of dots represent the $3\sigma$, $2\sigma$, and
  $1\sigma$ intervals of the experimental data. The yellow curve
  represents the distribution of $m_{\beta\!\beta}$. Detailed views of
  the densely distributed points are shown separately in the
  insets.
\label{picture3}
}
\end{figure}

The plots in figs.~\ref{picture1}, \ref{picture2} and \ref{picture3}
present the results of numerical scans.
The distributions of the neutrino masses for the cases 1, 2, and 3
look very similar to the distributions of the cases 5, 6, and 7
(according to the pattern of the Majorana phases), for
both normal and inverted hierarchies. Therefore in fig.~\ref{picture1}
we present the probability distributions of the neutrino masses only
for the first three cases.
Due to a small value of $\Delta m_{21}^2$ 
the distributions of the first and the second neutrino mass overlap in
most cases with an exception for the case 2 assuming normal hierarchy.

The oscillation angles $\sin^2\theta_{12}$ and $\sin^2\theta_{13}$
fill our scatter plots uniformly, however,
the values of $\sin^2\theta_{23}$ have specific probability
distributions in each case of table~\ref{tabl1}.
They are presented in fig.~\ref{picture2}
for normal and inverted hierarchies.

Another quantity that we can predict is the effective mass
$m_{\beta\!\beta}=|M_{ee}|$, which is related to the neutrinoless double
beta decay. Figure~\ref{picture3} shows the relations between the sum
of the neutrino masses and $|M_{ee}|$. The scatter plots for the cases
1, 2, and 3 look very similar to the plots of the cases 5, 6, and 7,
for both normal and inverted hierarchies. We present only
the first three cases.

%


\section{Summary} \label{summ}

The model with the symmetries $\mathbbm{Z}_2^{(\mathrm{int})}$ and
$CP$ fits well the phenomenological data for the lepton mixing angles
and for the neutrino mass-squared differences. The model predicts an
overall neutrino mass scale for normal and inverted neutrino mass
hierarchies.
When the Majorana phases are $e^{i\alpha_{21}}=+1$ and $e^{i\alpha_{31}}=-1$
(the cases 2 and 6 of table~\ref{tabl1}),
the neutrino mass scale is quite
low and fulfills the majority of the current cosmological bounds. The
Majorana phases have a greater influence to the studied distributions
than the Dirac phase. The
effective mass $m_{\beta\beta}$ has its median in the interval
$(0.01-0.33)$~eV, but the statistical majority of its values are
distributed in the range of $0.01-0.1$~eV.
The distribution
of $m_{\beta\!\beta}$ is different in each analysed case.


\acknowledgments

The authors thank the Lithuanian Academy of Sciences for the support
(project DaFi2015). D.J. thanks Luis Lavoura for the valuable
discussions and suggestions.


\newcommand{\hepth}[1]{\href{http://arxiv.org/abs/hep-th/#1}{\tt hep-th/#1}}
\newcommand{\hepph}[1]{\href{http://arxiv.org/abs/hep-ph/#1}{\tt hep-ph/#1}}
\newcommand{\nuclth}[1]{\href{http://arxiv.org/abs/nucl-th/#1}{\tt hep-ph/#1}}
\newcommand{\arXiv}[1]{\href{http://arxiv.org/abs/#1}{\tt arXiv:#1}}

\end{document}